\documentclass[prl,showpacs,twocolumn,floatfix]{revtex4}
\usepackage{graphicx}
\usepackage{amssymb}

\begin{document}
\title{Why humans die --- an unsolved biological problem}
\author{Mark Ya. Azbel'}
\affiliation{School of Physics and Astronomy, Tel-Aviv University, \\
Ramat Aviv, 69978 Tel Aviv, Israel}\thanks{Permanent address}
\affiliation{Max-Planck-Institute f\"{u}r Festkorperforschung -- CNRS, \\
F38042 Grenoble Cedex 9, France}

\begin{abstract}
Mortality is an instrument of natural selection. Evolutionary
motivated theories imply its irreversibility and life history
dependence. This is inconsistent with mortality data for
protected populations.  Accurate analysis yields mortality law,
which is specific for their  evolutionary unprecedented
conditions, yet universal for species as   evolutionary remote as
humans and flies. The law is exact, instantaneous, reversible,
stepwise,  and  allows for a rapid (within less than two years for
humans) and significant mortality decrease at any (but very old)
age.
\end{abstract}

\maketitle

  Mortality is an  instrument of natural selection. In the wild,
competition for sparce resources is fierce, and only relatively
few genetically fittest animals survive to their evolutionary
``goal'' -  reproduction. Even human life expectancy at birth was
around 40 - 45 years just a century ago (e.g., 38.64 years for
males in 1876 Switzerland \cite{database}). Evolutionary motivated
biological theories \cite{Charlesworth} imply that during and
beyond reproductive age mortality irreversibly increases and
strongly depends on the life history. This disagrees with the
demographic observation (see, e. g., \cite{Vaupel}) that mortality
is highly plastic; that different mortalities in Eastern and
Western Germanies converged few years after their unification;
that Norwegian females born in 1900 at 57 years reversed to
mortality they had 36 years younger (see later). Recent
experiments \cite{Mair} prove that the life-prolonging effect of
diet in fruit flies is independent of their past, starts
immediately and is lost when the dieting stops. Thus, in agreement
with theoretical predictions \cite{Azbel1}, mortality in protected
populations of species as remote as humans and flies has short
memory, is reversible and little depends on life history. Such
dominance of nurture over nature is inconsistent with any
evolutionary mechanism of mortality. But then, protected
populations are indeed evolutionary unprecedented. To unravel the
biologically unanticipated mechanism of their mortality, note that
similar situation triggered all breakthroughs in physics via
analysis of experimental data, which disregarded common wisdom and
preceded rather than followed any models. Luckily, accurate
forecasts of human mortality, and the resulting age structure of
the population, are important for economic, taxation, insurance,
etc, etc, purposes. That is why quantitative studies of mortality
were started \cite{Halley} long before Darvin, in 1693, by the
famous astronomer Halley (the discoverer of the Halley comet) and
followed in 1760 by the great mathematician Euler. To better
estimate and forecast mortality, demographers developed over 15
mortality approximations \cite{Coale}. Total mortality depends on
a multitude of unquantified factors which describe all kinds of
relevant details about the population and its environment, from
conception to the age of death: genotypes; life history; acquired
components; age specific factors; even the month of death and the
possibility of death being the late onset genetic decease
\cite{Charlesworth,Doblhammer}. Demographic approximations prove
that in a given country all these factors with remarkable accuracy
reduce to few parameters only. During the last century, mortality
rate in Western Europe at 0, 10 and 40 years decreased
correspondingly 50, 100, and 10 times. In contrast to such
mortality decrease (primarily due to improving living conditions,
medical ones included), the difference between mortality rates at
the same age in the same calendar year is rarely more than twofold
in different countries. However, demographers do not present
universal mortality approximations. They are interested in the
most accurate approximation of the most important and specific
mortality rate in a given country or its part. To a physicist
relative proximity of such approximations in different countries
suggests \cite{Azbel3} certain universality. Mortality in a
population is uncontrollably heterogeneous (e.g., $1891/1900$
female infant mortality is almost twice higher in Stockholm than
in the rural area \cite{stat}), and changes with time. It affects
different mortality characteristics in a different way.
``Additive'' variables, whose values in a heterogeneous population
are the averages over its different groups of the same age, are
invariant to any such averaging. The less heterogeneous the
population and its living conditions are, the more accurate their
mortality approximations are. I conjecture that a certain fraction
of mortality (denote it as ``canonic'') accurately yields the law,
which is the same (universal) in any population where
heterogeneity of additive mortality variables is restricted to
accurately quantified universal limits. Such universality is
sufficient to establish its law. Any heterogeneous population
consists of several ``restricted heterogeneity'' groups. Its
mortality reduces to the universal law and fractions of the
population in each of the restricted heterogeneity groups. Mortality in
different countries allows one to determine the universal law
parameters, and to verify its predictions.

  For males and females who died in a given country in a given
calendar year demographic life tables list, in particular,
``period'' probabilities $q(x)$ (for survivors to $x$) and $d(x)$
(for live newborns) to die between the ages
$x$ and $(x + 1)$ [note that $d(0) = q(0)$]; the probability
$l(x)$ to survive to $x$ for live newborns; the life expectancy
$e(x)$ at $x$. The tables also present \cite{database} the data
and procedure which allow one to calculate $q(x)$, $d(x)$, $l(x)$,
$e(x)$ for human cohorts, which were born in a given calendar
year.

Consider a heterogeneous population, consisting of the
groups with the number $N^{G}(x)$ of survivors to $x$ in a group
$G$. The total number of survivors $N(x)$ is the sum of $N^{G}(x)$
over all $G$. Thus, $l(x) = N(x) / N(0) = \sum c_{G} \ell^{G}(x) =
\langle \ell^{G}(x) \rangle$ is the average of $\ell^{G}(x)$ over
all groups, with $c_{G}$ and $\ell^{G}(x)$ being the ratio of the
population and the survivability to $x$ in the group $G$.
Similarly, $d(x) = \ell(x) - \ell(x + 1)$ reduces to its average
over the groups of the same age. In contrast, $q(x) = 1 - \ell(x +
1) / \ell(x)$ reduces to $q^{G}(y)$ in the groups of all ages $y$
from 0 to $x$, since the probability $\ell(x)$ to survive to $x$
equals $p(0) p(1) \ldots p(x - 1)$, where $p(y) = 1 - q(y)$ is the
probability to survive from $y$ to $(y + 1)$. The most age
specific additive variable is $d(x)$. (Naturally, it allows one to
calculate all other mortality characteristics, e. g., $q(x)$).
The most time specific one is $d(0) =
q(0)$ -- it depends on the time span less than 2 years (from
conception to 1 year). Thus, the most specific relation between
two additive variables is the relation between $d(x)$ and infant mortality
$q(0)$. A universal restriction on the heterogeneity of $q(0)$ is $q_{j} <
q^{G}(0) < q_{j + 1}$, where  $q_{j}$, $q_{j + 1}$ determine the
universal for all humans boundaries of the $j$-th interval.
Universal law implies that the relation between canonic $d(x)$ and
$q(0)$ in any universal interval is the same as the relation
between their values in any of the restricted heterogeneity
groups, i.e. that $d(x) = f_{x}[q(0)]$; $d^{G}(x) =
f_{x}[q^{G}(0)]$, where $f_{x}$ is a universal function. (Here and
further on $d$, $q$, etc denote canonic quantities). Since
additive $d(x) = \langle d^{G}(x) \rangle$, $q(0) = \langle
q^{G}(0) \rangle$, so $\left\langle f_{x}[q^{G}(0)] \right\rangle
= f_{x} \left[ \langle q^{G}(0) \rangle \right]$. According to a
simple property of stochastic variables \cite{Azbel3}, if the
average of a function is equal to the function of the average,
then the function is linear. So,
 \begin{equation} \label{eq1}
   d(x) = a_{j}(x) q(0) + b_{j}(x) \quad \text{if} \quad
     q_{j} < q(0) < q_{j + 1},
 \end{equation}
where parameters $a_{j}$, $b_{j}$ for a given $x$ are universal
constants. (Here and on I skip their argument $x$).  When canonic
infant mortality $q(0)$ reaches an interval boundary (\ref{eq1}),
it must homogenize to the boundary value. Since $d(x)$ at all ages
reduce to infant mortality, they simultaneously reach the interval
boundary and, together with $q(0)$, homogenize there. [Two such
``ultimate'' boundaries  are well known:  $q(x) = 0$ when nobody
dies, and  $q(x) = 1$ when nobody survives at the age $x$]. At
different intervals linear relations are different. Thus, the
universal law implies piecewise linear $d(x)$ vs $q(0)$ with
simultaneous at all ages intersections. Any heterogeneous
population is distributed at the universal intervals. Their
occupation and Eq. (\ref{eq1}) determine piecewise linear, but
non-universal relation between $d(x)$ and $d(0) = q(0)$. At a
given non-universal linear segment
 \begin{equation} \label{eq2}
   d(x) = aq(0) + b
 \end{equation}
For a given $x$ non-universal $a$ and $b$ reduce to the universal
law parameters and to the non-universal fractions of the
population in each of its linear intervals. Heterogeneity of
mortality in some groups in a given country may be sufficiently
low to fit into a single universal interval. Then they yield the
universal law and allow for a comprehensive study of canonic
mortality. Suppose that a heterogeneous population is distributed
at two, e.g., the 1-st and 2-nd, universal intervals with the
concentrations $c_{1}$ and $c_{2} = 1 - c_{1}$ correspondingly.
Then  $q(0) = c_{1} q_{1}(0) + (1 - c_{1}) q_{2}(0)$; $d(x) =
c_{1} d_{1}(x) + (1 - c_{1}) d_{2}(x)$. Thus, by Eq. (\ref{eq1},
\ref{eq2}), $q_{1}(0) = \alpha_{1} q(0)$, $q_{2}(0) = \alpha_{2}
q(0)$, where $c_{1} = (b_{2} - b) / (b_{2} - b_{1})$, and
$\alpha_{1} = (a - a_{2}) / [c_{1}(a_{1} - a_{2})]$; $\alpha_{2} =
(a_{1} - a) / [(1 - c_{1})(a_{1} - a_{2})]$. The crossover to the
next non-universal segment occurs when, e.g., $q_{1}(0)$ reaches
the intersection $q^{U}(0) = (b_{2} - b_{1}) / (a_{1} - a_{2})$ of
the first and second universal segments. Then  $q_{1}(0) =
q^{U}(0)$ implies $d^{I}(x) = a_{1} q^{I}(x) + b_{1}$. (A
subscript $I$ denotes an intersection). So, this non-universal
intersection belongs to the first universal linear segment or its
extension. Inversely, such universality is the criterion of the
population distributed at two universal segments. The set of such
``extended'' universal segments yields the universal law, while
its non-universal intersections determine the fractions of the
populations at the universal intervals. A general case (when the
population is distributed at more than two universal segments) is
more complicated, but also reduces to the universal law and the
population fractions at the segments.

  To quantitatively verify the universal law, consider the number
$D(x)$ of, e.g., Swiss female deaths at a given age $x$ in each
calendar year from 1876 to 2001\cite{database}. At 10 years $D(10)$
decreases (together
with mortality) from 126 in 1876 to 1 in 2001. At 80 years $D(80)$
increases (together with the life expectancy) from $231$
to $951$. It depends on the size of the population, e.g. in
1999 Japan at 80 years it is $13, 061$. Its stochastic error
is $\sim 1 / D^{1/2}$. If demographic fluctuations in mortality
are consistent with this (minimal for a stochastic quantity)
generic error for the same age, denote the corresponding mortality
as ``regular''. Otherwise, denote it as ``irregular''. Remarkably,
mortality is irregular only during, and few years after, major
wars, epidemics, food and water contamination, etc., when its
change within few years is not relatively small.

\begin{figure}[tb]
  \includegraphics[clip=true,width=8cm]{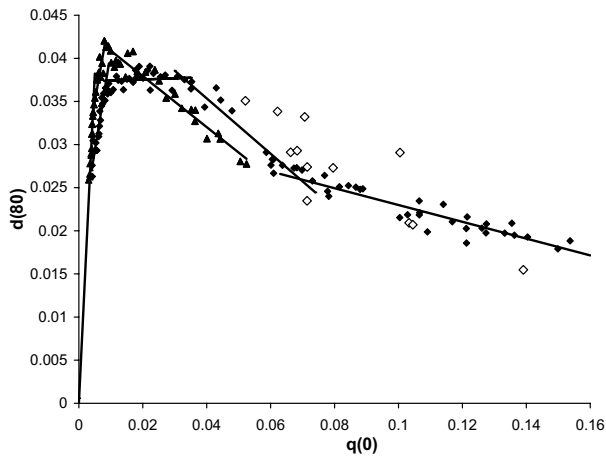}
  \caption{The probability $d(80)$ for newborn females in 1898-2001
  France (diamonds) and 1950-1999 Japan (triangles) to die between
  80 and 81 years of age vs the same calendar year infant mortality
  $q(0)$. Empty diamonds correspond to 1918 flu pandemic and World
  Wars. They are disregarded in the linear regressions (straight
  lines), which minimize the mean linear deviations from black signs
  to statistical $5 \%$. When Japanese $q(0) = 0$, the corresponding
  linear regression yields $d(80) = 0$.}\label{fig1}
\end{figure}

\begin{figure}[tb]
  \includegraphics[clip=true,width=8cm]{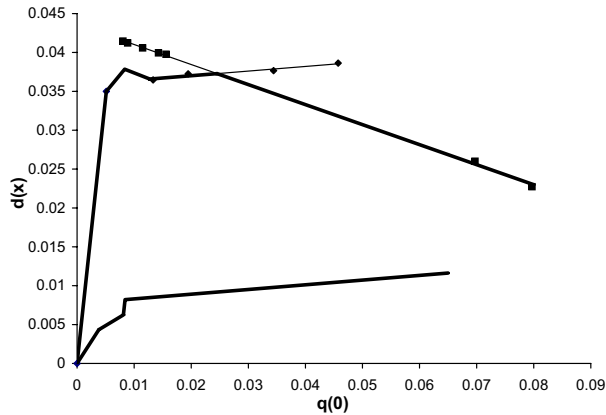}
  \caption{Universal law of the canonic $d(80)$ and $d(60)$ (upper
  and lower curves, thick lines) vs $q(0)$. Note that both
  $d(80) = 0$ and $d(60) = 0$ when $q(0) = 0$. Diamonds and squares
  represent non-universal intersections for (from left to right)
  England (two successive intersections), France, Italy and Japan,
  Finland, Netherland, Norway, Denmark, France, England
  correspondingly. Thin lines extend the universal linear segments.}
  \label{fig2}
\end{figure}

To verify and determine the universal law,
I approximate regular empirical $d(x)$ vs $d(0)$  with the minimal
number of linear segments which yields their statistical accuracy
for each given age $x$. [Figure \ref{fig1}
presents the examples of $d(80)$ vs. $q(0)$ for Japanese and
French females.] Demographic data demonstrate that all
non-universal intersections in most developed countries fall on
the universal straight lines (see examples in Fig. \ref{fig2}).
Thus, the population in each such country reduces to 2 restricted
heterogeneity groups (which change at the non-universal
intersections). Then non-universal intersections determine
the universal law (see Fig.
\ref{fig2}). The law may be refined by accounting for a
(relatively small) contribution from more than two restricted
heterogeneity groups.

\begin{figure}[tb]
  \includegraphics[clip=true,width=8cm]{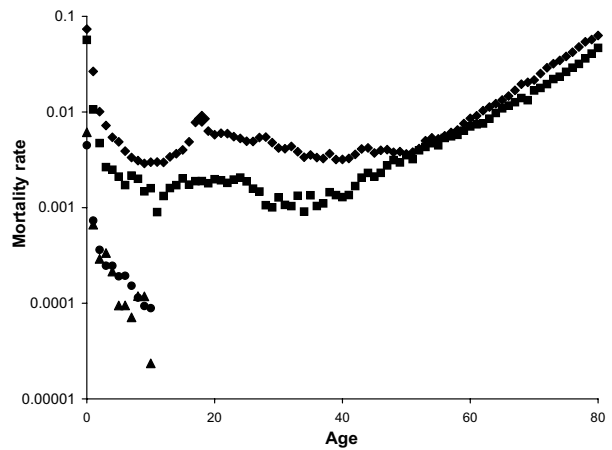}
  \caption{Mortality rates vs age in Norwegian 1900 (diamonds, the
  large one for 1918 year of flu pandemic in Europe), Swedish 1920
  (squares), Japanese 1989 (circles), Swiss 1990 (triangles)
  female cohorts.}\label{fig3}
\end{figure}

  Consider other implications and predictions of the universal law.
The extrapolation of the Japanese piecewise linear dependence in
Fig. \ref{fig1} to $q(0) = 0$ yields $d(80) = 0$, i.e. zero
mortality at (and presumably prior to) 80 years. Similarly,
$d(60)$ and $d(80)$ in Fig. \ref{fig2} universally $\rightarrow 0$
when $q(0) \rightarrow 0$. This is consistent with, e.g., the
dependencies of the life expectancies $e(0)$ at birth and $e(80)$
at 80 years on the values of the same calendar year birth
mortality $q(0)$ for Japanese and French females. If nobody dies
until 80, then $e(0) = 80 + e(80)$. In fact, the extrapolated
$e(0) = 93$ years,  $e(80) = 16$. Thus, $e(80) + 80 = 96$ years is
just $3 \%$ higher than $e(0)$. Vanishing mortality may have
already been observed. In 2001 Switzerland less than 17 females
died in any age group from 1 till 26 years,  43 died at 40
years. In Japan, where the population is 18 times larger, 50 girls
died at 10 years in 1999 (cf. Fig. \ref{fig3}). Such values of a
stochastic quantity are consistent with zero mortality in the
lowest mortality groups.

  The universal law reduces the period canonic mortality at any age
to the infant mortality $q(0)$. So, together with $q(0)$, at any
age regular mortality rate may be rapidly reduced and reversed to
its value at a much younger age. This agrees with the mortality of
Norwegian and Swedish female cohorts, born in 1900 and 1920
correspondingly (Fig. \ref{fig3}).
Both countries were neutral in the World Wars. In both infant
mortality $q(0)$ is higher than mortality $q(80)$ at 80 years. In
Sweden $q(0)$ decreases 63-fold to $q(11)$, then increases
2.3-fold to $q(24)$. Thereafter it decreases to the same value at
34 years as 23 years earlier, at 11, and only at 45 years reaches
almost the same value as 30 years younger, at 15. In Norway $q(0)$
decreases 24.5-fold to $q(9)$, doubles at $x = 21$, halves back to
the minimal value at $x = 34$, and  then slowly changes, until at
57 years it restores the mortality it had at 21, i.e. 36 years
younger \cite{Azbel4}.

  Rapid crossovers in mortality (see Fig. \ref{fig1}) expose
several modes of the universal mechanism, which switch
simultaneously for all ages (see Fig. \ref{fig2}). The changes
amplify significant declines of old age mortality in the second
half of the 20-th century \cite{Wilmoth}, with its spectacular
medical progress. Predicted homogenization of the mortality at the
intersections was noticed for male and female mortality
\cite{Azbel1}.

  The non-universal law determines the non-universal population
fractions in different universal intervals for a given country,
sex, and calendar year. The fractions depend on genetics, life
history, mutation accumulation and other factors. The difference
between the total mortality and its piecewise linear approximation
may be partitioned into stochastic fluctuations (which yield the
Gaussian distribution), singular ``irregular fluctuations''
(related to, e.g., 1918 flu pandemic and world wars), and
systematic deviations (related to evolutionary mechanisms
\cite{Charlesworth, Azbel1} and all unaccounted for factors).
Depending on age, from 3 till 95 years the latter contribute
from $2 \%$ to $10 \%$ of the total mortality. The approach may be
refined if one considers several additive parameters. Preliminary
study suggests this little increases the accuracy.

  Regular mortality also dominates in protected populations of
flies. The relations between their additive mortality variables
are also piecewise linear \cite{Azbel1}. Thus, their mortality is
also predominantly universal. The (properly scaled) law, which is
universal for species as remote as humans and well
protected populations of flies \cite{Azbel1} despite their different evolution,
may be considered a conservation law in biology and evolution. One
wonders how, why and when a law, which is specific for
evolutionary unprecedented (protected) populations, could
evolutionary emerge. It suggests a possibility of similar
``evolutionary unmotivated'' laws for other biological
characteristics.

  To summarize. Universal law of mortality specifies the groups
whose infant mortality heterogeneity is restricted by the
universal limits. In any population of any age, the law
rapidly (on the scale of two
years for humans) adjusts a dominant fraction of the
total mortality to infant mortality and the fractions of the
population in these groups. This implies that, in contrast to
mortality in the wild, mortality in well protected populations is
dominated by a genetically programmed intrinsic mechanism which
provides its unprecedented rapid adaptation to current living
conditions. The universal form and relatively high accuracy
(mostly on the scale of mortality fluctuations) of the law make
it a universal and biologically explicit demographic approximation
of the total mortality. The law provides certain clues to its
possible physiology and physical model.


\end{document}